\documentclass[a4paper,11pt]{article}
\usepackage{booktabs}
\usepackage{bm}
\usepackage{float}
\usepackage{latexsym}
\usepackage{dcolumn}
\usepackage{amsfonts,amssymb}
\usepackage{graphicx}
\usepackage{epsfig}
\usepackage{psfrag}
\usepackage{nccmath}
\usepackage{moresize}
\usepackage{enumerate}
\usepackage[hang,flushmargin]{footmisc}
\usepackage[titletoc,toc]{appendix}
\usepackage{lipsum}
\usepackage{subcaption}
\usepackage{hyperref}
\usepackage{rotating}
\usepackage{multirow}
\usepackage{multicol}
\usepackage{xfrac}
\usepackage{mathtools}
\usepackage{nccmath}
\usepackage{cite}
\usepackage{placeins}
\usepackage{tikz}
\usepackage{textalpha}
\usepackage[font=small,labelfont=bf]{caption}
\usepackage{wrapfig}
\usepackage{authblk}
\usepackage{cite}
\usepackage[utf8]{inputenc}
\usepackage[T1]{fontenc}
\usepackage{enumitem}
\usepackage{blindtext}

\usetikzlibrary{positioning}
\graphicspath{{figures/}}

\begin{document}

\title{{\LARGE The Oscillatory Universe, phantom crossing and the Hubble
tension}}
\author{{\large Mohit K. Sharma}$^{1}$ {\large , Shibesh Kumar Jas Pacif}$%
^{2}$ {\large , Gulmira Yergaliyeva}$^{3}$ {\large , Kuralay Yesmakhanova}$%
^{4}$ \\
$^{1,2}$Centre for Cosmology and Science Popularization (CCSP), SGT
University, Delhi-NCR, Gurugram 122505, Haryana, \ India \\
$^{3,4}$Eurasian International Center for Theoretical Physics, Eurasian
National University, Nur-Sultan, 010008, \ Kazakhstan \\
Email: $^{1}$ mr.mohit254@gmail.com , $^{2}$shibesh.math@gmail.com}
\date{}
\maketitle

\begin{abstract}
We investigate the validity of cosmological models with an oscillating scale
factor in relation to late-time cosmological observations. We show that
these models not only meet the required late time observational constraints,
but can also alleviate the Hubble tension. As a generic feature of the
model, the Hubble parameter increases near the current epoch due to its
cyclical nature exhibiting the phantom nature allowing to address the said
issue related to late time acceleration.
\end{abstract}

\section{Introduction}

\label{intro} One of the fundamental concerns of cosmology is the late-time
accelerated expansion of the universe \cite{Riess1998, Perlmutter1999},
which is possibly driven by the presence of a Dark energy (DE) component.
Even if the $\Lambda $CDM ($\Lambda $ is the Cosmological constant and CDM
is the Cold Dark Matter) is statistically most supported by observations,
the fine-tuning and coincidence problem associated with it motivates one to
look for alternative theories of gravity \cite{DEbook, Copeland, Planck18}. As we know that
cosmological observations are in good agreement with the $\Lambda $CDM
model with high precision, they still allow a narrow but sufficient room for the existence of weakly dynamical DE cosmological models. 
In this context, a variety of scalar fields inspired by high energy physics 
and phenomenological viewpoints have been examined in the literature to 
describe the late time acceleration \cite{sami_1,sami_2,Copeland}.

Apart from being well-suited, atleast, for the high-redshift measurements, $%
\Lambda $CDM is confronted with the low-redshift data which measures a
larger value of the Hubble constant $H_{0}$. In particular, the
observational value of $H_{_{0}}$ found by Planck satellite is $67.36\pm
0.54 $ Km/s/Mpc, whereas, low-redshift surveys, such as the
Supernovae $H_{_{0}}$ for \textbf{the Equation of State (SH0ES)
collaboration finds }$H_{_{0}}=73.04\pm 1.04$\textbf{\ Km/s/Mpc \cite{SH0ES}%
, Megamaser Cosmology Project finds $73.9 \pm 3.0$ Km/s/Mpc, and 
H0LiCOW finds $73.3^{+1.7}_{-1.8}$ Km/s/Mpc}. This large discrepancy between high and low-redshift measurements of $%
H_{_{0}}$ which is up to $4.2\sigma $ level has given a lot of attention to
alternative cosmological models which even if cannot completely alleviate
this `tension' completely, at least try to reduce it to some extent 
\cite{KHOURY2016, Hub5, Hub10, mohit, HTen1,HTen2, HTen3, HTen4, HT1, HT2, HT3}.

Despite the existence of several scalar field DE models or modified gravity
theories that might ease this tension, the problem can also be seen from a
completely other viewpoint that takes into account the universe's
oscillatory characteristic. The Quasi Steady State Cosmology (QSSC), in
instance, is one such domain of cosmology that permits the universe's scale
factor to oscillate rather than increase monotonically. It is an alternative
cosmological theory developed in \cite{qssc1} and further discussed in \cite%
{qssc2}, \cite{qssc3}. In the literature, the theoretical framework and
exact solutions \cite{qssc4, qssc5}, light nuclei production \cite{qssc6,
qssc7, qssc8}, structure formation \cite{qssc-nayeri}, some observational
tests \cite{qssc9, qssc10} and stability of the model \cite{qssc11} have
already been discussed. In recent times, QSSC scenarios have also been
explored with respect to the late-time observations \cite{qssc-thobs,
qssc12, qssc13, qssc14, qssc15, qssc16, qssc17, qssc18}. One of the
significant feature of QSSC is that the model is free from initial
singularity problem and also capable of explaining the accelerated expansion
of the Universe. So, it is interesting to study QSSC to look for some more
scope to resolve some fundamental problems. Here, in this paper, we are
intended to work on the Hubble tension and see the possibility of resolving
Hubble tension by QSSC. One of the most important features of QSSC 
scenarios is that they avoid any Big-Bang singularities while still explaining
late-time cosmic acceleration and their potential to solve the Hubble
tension problem.

Our major goal in this paper is to investigate the late-time cosmological
evolution using QSSC and to determine the amount of Hubble tension relieved
owing to the oscillatory character of the Hubble parameter $H(t)$. QSSC is
effective of tracing back the evolutionary history of the universe at least
at the background level up to large redshifts without defining any type of
fluid or field. Fluid with equation of state parameter $w_{x}\gtrsim -1$
always gives rise to $\dot{H}\lesssim 0$ (where $^{\cdot }\equiv d/dt$, and $%
t$ is the cosmic time), therefore in order to have a supper acceleration $%
\dot{H}\gtrsim 0$ one requires $w_{x}\lesssim -1$. It was demonstrated in
Ref \cite{ref1}, that one needs at least two fields for consistent
realization of the late-time enhancement in the Hubble parameter. Indeed,
one can formally construct single scalar field which gives rise to
transition across the phantom divide \cite{ref2}. However, such transitions
are physically implausible in single field models as they are either
realized by a discrete set of trajectories in the phase space or are
unstable under cosmological perturbations. To describe the said behaviour
consistently, one may use two field framework dubbed quintom models \cite%
{quint1, quint2, quint3}. \textbf{However, it is possible to realize }$%
w_{x}\lesssim -1$\textbf{\ with one scalar field in scalar-tensor gravity
e.g. In Ref \cite{ADD1, ADD2}, the reconstruction of expansion law in
modified gravity is discussed.}

The layout of our paper is as follows: The section \ref{intro} provides a
short introduction on dark energy and Hubble tension. In section \ref%
{background}, in a typical flat-Friedmann universe, we first present a brief
cosmic background description of the QSSC model, then in section \ref%
{methodology}, we describe the technical details of our statistical analysis
done using the Monte Carlo Markov Chain (MCMC) simulations using
observational datasets. Finally, a summary and the conclusions of this work
are given in section~\ref{conclusion}.

\section{Background level framework of QSSC}

\label{background}

Let us begin with a consideration that the oscillatory universe satisfies
the conditions of homogeneity and isotropy and thus can be described by the
standard FRW metric: $ds^{2}=dt^{2}-a(t)^{2}dr^{2}$ ($t$:= comoving time, $%
a(t)$:= scale factor, and $r$:= comoving distance). Unlike the common
perspective of the universe in the standard FRW universe, which holds that
it only expands with time, the QSSC model provides a more generalised view
to this. This not only supports our universe's traditional evolutionary
paradigm (at least in the long run), but it also solves the challenges
related with singularities. In fact, the crucial aspect to consider QSSC
model lies in its ability to give rise to a cosmological scenario which
allows our Universe to not just keep expanding but also to contract in
different cycles of its evolution. Because the scale factor encodes our
Universe's time-evolutionary profile, one might suggest the following form
of the scale factor to actualize such an oscillatory universe: 
\begin{equation}
a(t)=\exp \left( \frac{t}{P}\right) \left[ 1+\eta \cos \frac{2\pi t}{Q}%
\right] \,,  \label{a(t)}
\end{equation}%
where $P$ and $Q$ have dimensions of time, and $\eta $ is a dimensionless
constant. This choice of the form of the scale factor is well described in a
series of papers by Hoyle, Burbidge and Narlikar \cite{qssc1}, \cite{qssc2}, 
\cite{qssc3}, \cite{qssc4}, \cite{qssc5}. The main essence in this
parametrization of $a(t)$ lies in follows:

\begin{itemize}
\item In the limit $\eta \rightarrow 0$ it gives rise to de-Sitter like
evolution.

\item In the limit $t\rightarrow 0$, $a(t)\rightarrow 1+\eta $. Thus, the
standard Big-Bang singularity does not occur in QSSC scenarios.
\end{itemize}

As $P$ and $Q$ are both dimensional quantities, let us make following
propositions: 
\begin{equation}
P=\alpha t_{0}\,,\quad Q=\beta t_{0}\,,  \label{P-Q}
\end{equation}%
where $\alpha $ and $\beta $, by definitions, are dimensionless parameters,
and $t_{0}$ is the present value of the cosmic time $t$.

Consequently, the Hubble parameter $H(t)$ from Eq.\,(\ref{a(t)}) is
expressed as 
\begin{equation}  \label{H}
H(t):=\frac{\dot{a}(t)}{a(t)}=\frac{1}{\alpha t_{0}}-\frac{2\pi \eta \sin
(\zeta _{t})}{\beta \left[ 1+\eta \cos (\zeta _{t})\right] t_{0}}\,,\qquad %
\mbox{where}\quad \zeta _{t}=\frac{2\pi t}{\beta t_{0}}\,.
\end{equation}
Note that when $\eta \to 0$, the oscillatory behavior of the universe ceases
to exist and the Hubble parameter becomes constant. 
\begin{figure}[htbp]
\centering
\begin{subfigure}{\textwidth}
    \includegraphics[width=6.5in, height=2in]{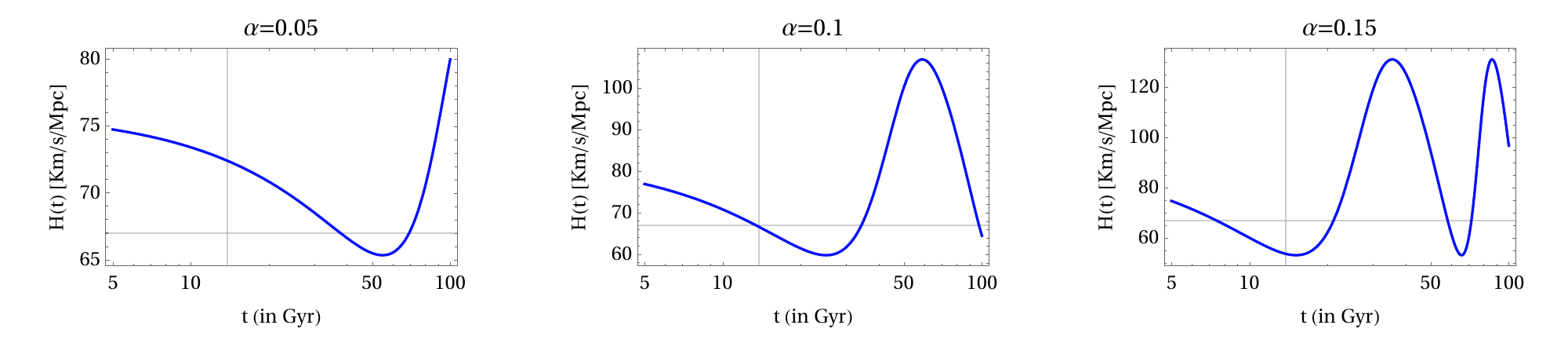}
    \caption{For $\beta=0.7$ and $\eta=0.3$}
    \label{fig:demo1}
  \end{subfigure}
\begin{subfigure}{\textwidth}
    \includegraphics[width=6.5in, height=2in]{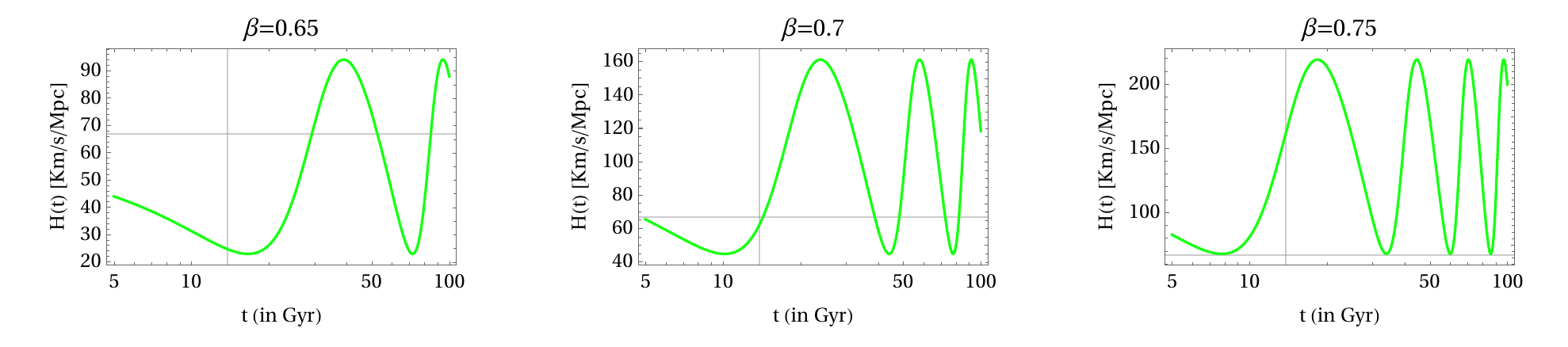}
    \caption{For $\alpha=0.2$ and $\eta=0.3$}
    \label{fig:demo2}
  \end{subfigure}
\begin{subfigure}{\textwidth}
    \includegraphics[width=6.5in, height=2in]{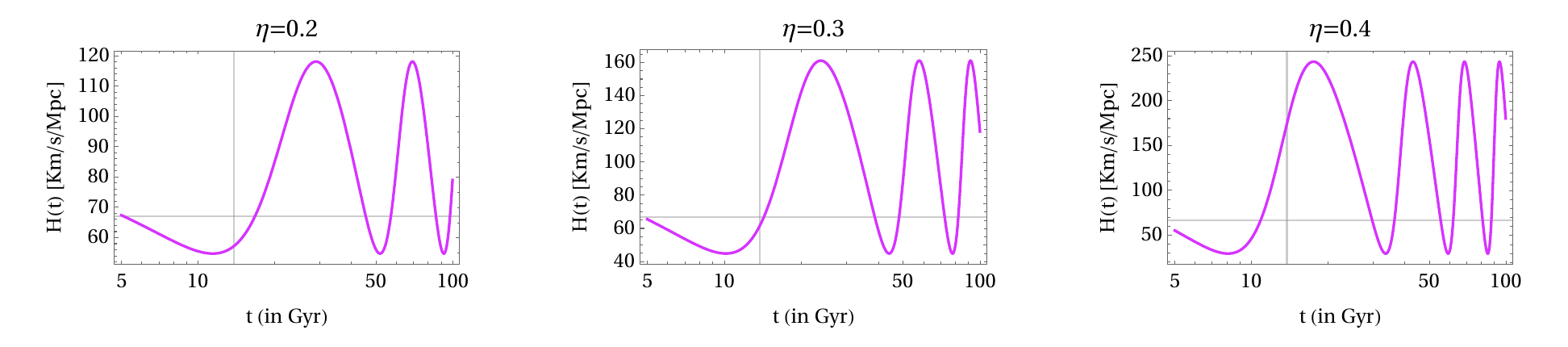}
    \caption{For $\alpha=0.2$ and $\beta=0.7$}
    \label{fig:demo3}
  \end{subfigure}
\caption{Evolution of $H(t)$ with cosmic time $t$ for different set of
fiducial values of $\protect\alpha$, $\protect\beta$ and $\protect\eta$. The
horizontal and vertical lines represents the observed value of the Hubble
constant (i.e. $67$\, Km/s/Mpc) and the age of the universe, respectively.}
\end{figure}

From the above equation, one can extract out the present value of cosmic
time i.e. $t_0$ in terms of $H_0$ as 
\begin{equation}
t_{0}=\frac{1}{\alpha H_{0}}-\frac{2\pi \eta \sin (\zeta _{0})}{\beta H_{0}%
\left[ 1+\eta \cos (\zeta _{0})\right] }\,,\qquad \mbox{where}\quad \zeta
_{0}=\frac{2\pi }{\beta }\,.  \label{t_0}
\end{equation}
One may note that in order to realize a standard non-oscillatory Universe, $%
\eta $ must vanish which also make the cosmological evolution independent of 
$\beta $. As a result, the frequency of oscillations due to $\beta $ is
controlled by parameter $\eta $. It is also interesting to note that the
remaining parameter $\alpha $, then, can be treated as the power of time in
the standard cosmological evolutionary profiles of $a(t)$, such as $1/2$ and 
$2/3$ in radiation and matter dominated era, respectively.

\begin{figure}[!ht]
\centering
\begin{subfigure}{0.495\linewidth} \centering 
   \includegraphics[height=6cm,width=6.4cm]{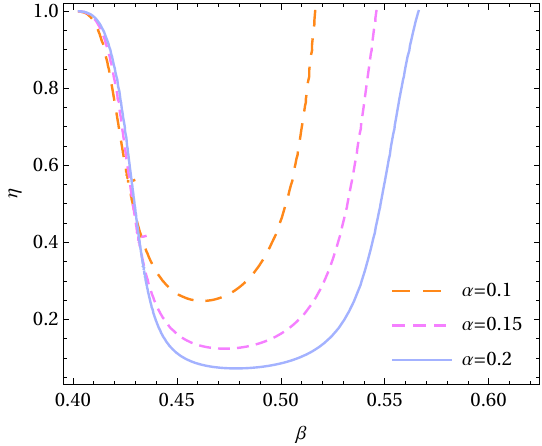}
   \caption{For fiducial values of $\alpha$} \label{LCDM-a}
\end{subfigure}
\begin{subfigure}{0.495\linewidth} \centering
    \includegraphics[height=6cm,width=6.4cm]{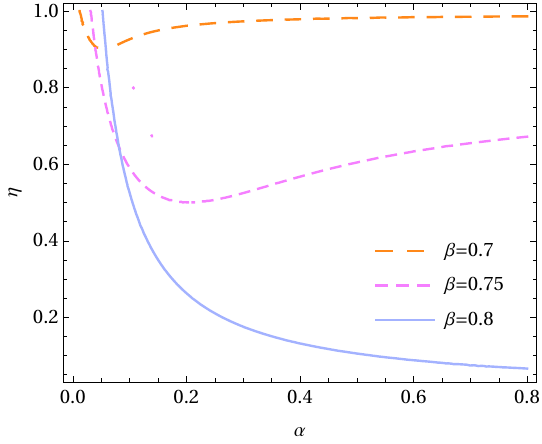}
    \caption{For fiducial values of $\beta$} \label{LCDM-b}
\end{subfigure}
\caption{Contours of figs. (a) and (b) represent cases when the QSSC model
reduces to the $\Lambda$CDM model with DE equation of state equal to $-1$.}
\label{fig_lcdm}
\end{figure}

In order to find the $\Lambda$CDM limit(s) of QSSC model, we first identify
the total equation of state parameter $w_{_{tot}}$ in the latter, which is
defined as 
\begin{equation}  \label{wtotal}
w_{_{tot}}(t) = -1- \frac{2}{3}\,\frac{\dot{H}(t)}{H(t)^2} \, ,
\end{equation}
such that $\dot{H}(t)$ is given by 
\begin{equation}  \label{Hdot}
\dot{H}(t) = \frac{4H^2 \pi^2 \alpha^2 \eta (1+\eta \cos(\zeta_0))^2 
\mathcal{X}(t)}{\mathcal{Y}(t)^2\left[\beta(1+\eta \cos \zeta_0)-2\pi \alpha
\eta \sin(\zeta_0)\right]^2} \, ,
\end{equation}
where 
\begin{equation}
\mathcal{X}(t)=\eta+\cos \left(\frac{2H\pi \, t}{\frac{\beta}{\alpha}- \frac{%
2\pi \eta \sin(\zeta_0)}{1+\eta \cos(\zeta_0)}}\right) \, , \quad \mathcal{Y}%
(t) = 1+\eta \cos \left(\frac{2H\pi \, t}{\frac{\beta}{\alpha}- \frac{2\pi
\eta \sin(\zeta_0)}{1+\eta \cos(\zeta_0)}}\right) \, .  \nonumber
\end{equation}
Using Eqs. (\ref{H}), (\ref{wtotal}) and (\ref{Hdot}), one can calculate the
evolution of $w_{_{tot}}$ for a given parametric values of model parameters.
In order to estimate the effective DE equation of state parameter for QSSC
model, let us make use of the total equation of state expression i.e. 
\begin{equation}  \label{w_param}
w_{_{tot}} = w_{_{DE}} \Omega_{_{DE}} + w_{_{M}} \Omega_{_{M}}
\end{equation}
where $\Omega_{_{M}}$ is the matter density parameter. Since, matter is
considered to be pressure-less, therefore, one can take $w_{_{M}}=0$. Due to
the numerical complications involve in QSSC to extract out the exact profile
of DE equation of state, we therefore restrict ourselves to the present
epoch ($z=0$) only. By combining Eqs. (\ref{wtotal}) and (\ref{w_param}),
and taking fiducial values $H_0=70$ Km/s/Mpc and $\Omega_{_{DE}}=0.67$, we
can determine our parametric ranges which can give us the $\Lambda$CDM limit
i.e. $w_{_{DE}}(0)=-1$. In figs. (\ref{fig_lcdm}) (a) and (b) we depict the
contour lines for different values of $\alpha$ and $\beta$ which corresponds
to the $\Lambda$CDM case at the present epoch.

In order to estimate our set of four parameters $p=\{ \alpha, \beta, \eta, h
\}$ ($h=H_0/(100 \, km/s/Mpc)$), one has to express Eq.\,(\ref{H}) in terms
of redshift $z$ or scale factor $a(t)$. However, due to the complicated
expression of Eq. (\ref{a(t)}), exact analytical form of $t$ in terms of $%
a(t)$ or $z$ is difficult to obtain. Hence, we numerically evaluate $t$ from
Eq. (\ref{a(t)}) and then put it back in Eq. (\ref{H}). In this way, one can
evaluate $H(z)$ which can be used for parametric estimations using
observational datasets. However, as we see that the scale factor $a(t)$ is
not normalized at the present time $t_0$, hence the redshift $z$ is related
to scale factor as 
\begin{equation}  \label{redshift}
1+z = \frac{a(t_0)}{a(t)} \, , \quad \mbox{where} \quad a(t_0) = \exp\left(%
\frac{1}{\alpha}\right)\left[1+\eta \cos (\zeta_0) \right] \, .
\end{equation}
One may also note that in order to be compatible with observations which
although does not signal any kind of oscillatory behaviour of the universe
atleast in the past, the oscillatory behaviour of the scale factor will be
tightly constrained. In fact, the parameter $\eta$ must be less than unity
and its any non-zero value will cause the oscillations. Also, the time
period of oscillations should be equal to or greater than the age of the
universe otherwise it can affect several stringent constraints imposed by
CMB observations, type Ia supernovae, LSS, etc..

With above formulation, we can now proceed to perform our data analysis in
the next section.

\section{Observational constraints on QSSC}

\label{methodology}

For our parametric estimations, we perform the Markov Chain Monte Carlo
simulation on our model using the $H(z)$+Pantheon(binned)+BAO dataset. In
order to do that, we define the Likelihood function $\mathcal{L}$ as 
\begin{equation}
	\mathcal{L}\propto \exp \left( -\frac{\chi ^{2}}{2}\right) \,.
	\label{Likelihood}
\end{equation}

In our parametric estimations, we take a set of thirty $H(z)$ measurements
in the range $z\in \lbrack 0.07,1.965]$ given in \cite{Morescoetal2016}. 
\textbf{For the Supernova Type-Ia Pantheon, we take Hubble rate $E(z_i) = H(z_i)/H_0$ data points for six different redshifts within the range $z=0.07$ to $z=1.5$ presented by Reiss et al in \cite{riess}. This effectively compressing the informations of the - Pantheon compilation \cite{pantheon} and 15 SnIa at 
	$z >1$ of the CANDELS and CLASH Multy-Cycle Treasury (MCT) programs obtained by the HST, nine of which are at $1.5 <z <2.3$. We have also omitted the data point at $z=1.5$ which does not follow a Gaussian distribution (see \cite{amendola} for more 
	details)}. In addition, we take two points for BAO Lyman-$\alpha 
$ \footnote{%
	Note that since the model under consideration does not include any explicit
	form of present-day matter density parameter, therefore, the implementation
	of other BAO dataset becomes cumbersome. This is due to the fact that $%
	\alpha $, $\beta $ intrinsically encoded the the information for matter
	density parameter and therefore, cannot constrain the matter density
	parameter separately}.

Hence, for combined dataset, the individual $\chi ^{2}$ gets added to give 
\begin{equation}
\chi _{total}^{2}=\chi _{OHD}^{2}+\chi _{Pan}^{2}+\chi _{BAO}^{2}\,.
\label{total-chi}
\end{equation}

\begin{figure}[tbp]
\centering
\includegraphics[height=12cm, width=12cm]{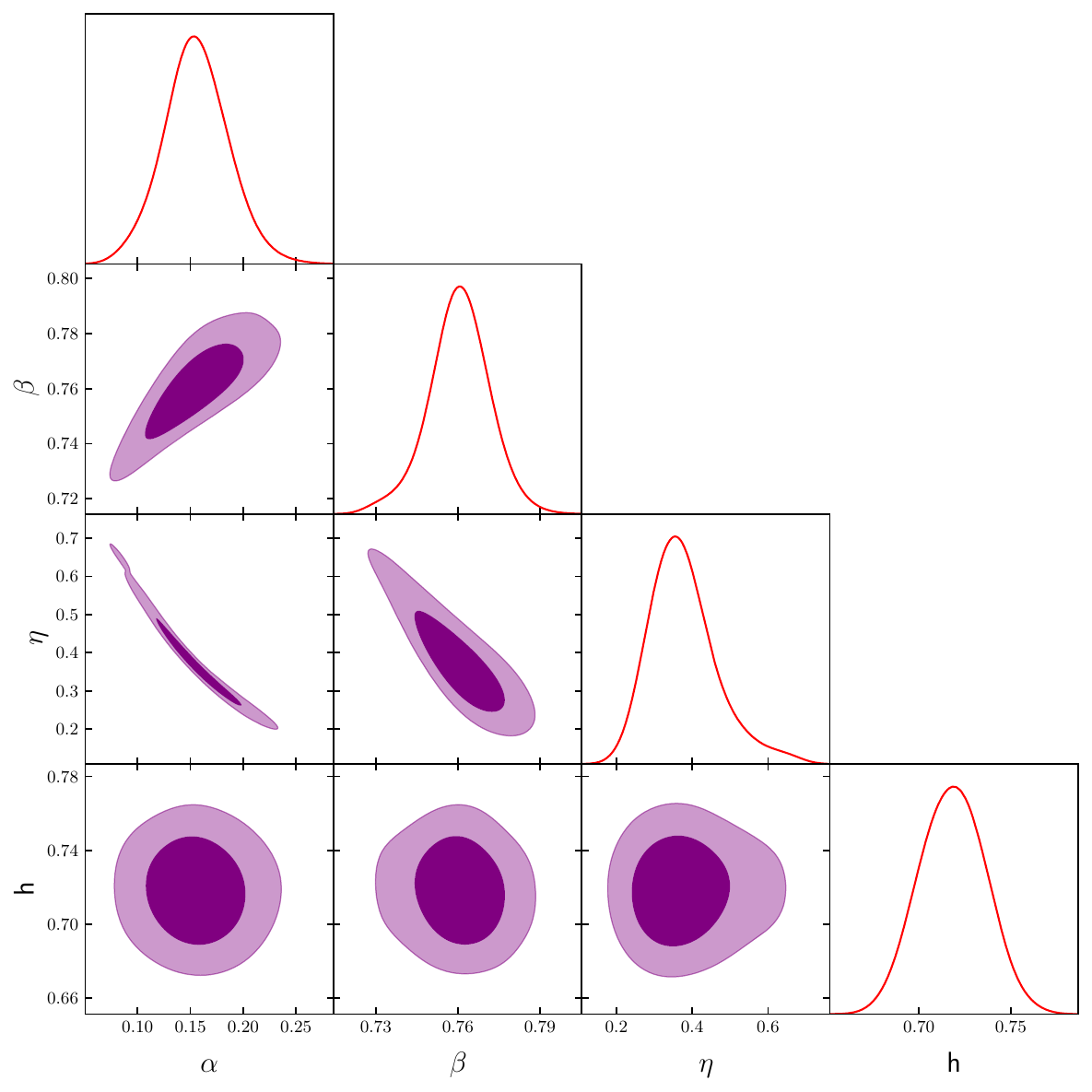}
\caption{Upto $2\protect\sigma$ parametric regions between $\protect\alpha$, 
$\protect\beta$, $\protect\eta$ and $h$ for combined OHD+Pantheon+BAO
dataset. }
\label{fig:contour}
\end{figure}

\textbf{Though the }$SH0ES$\textbf{\ Collaboration measured the SN absolute
magnitude directly, the difference may be less significant for late-time
adjustments, such as changing the universe's expansion pace, than it may be
for early-time modifications \cite{AddData1, AddData2}. However, we make an
effort to fit our model for a Gaussian prior to the Hubble constant
measurement provided by the }$SH0ES$\textbf{\ Collaboration and in order to
run our MCMC simulation, we give the following prior range for our parameters%
}: 
\begin{table*}[tbh]
\centering
\begin{tabular}{|c|c|}
\hline
Parameters & Priors \\ \hline
$\alpha$ & [-0.3 , 0.3 ] \\ \hline
$\beta$ & [0.5, 1] \\ \hline
$\eta$ & [0, 0.5] \\ \hline
$h$ & [0.55 ,0.85 ] \\ \hline
\end{tabular}%
\caption{Priors for the MCMC parameters}
\label{priorTab}
\end{table*}
The obtained $2\sigma $ parametric contours between our set of parameters $%
\alpha $, $\beta $, $\eta $, $h$ are shown in fig. (\ref{fig:contour}). In
that plot one can see that the allowed parametric region for combined
dataset is, as expected, smaller than that of the OHD. Also, one observes
high correlation of $\eta $ with both $\alpha $ and $\beta $. The estimated
values of our parameters are enlisted below. It is interesting to note that
the estimated value of the Hubble constant $h$ tends to be higher than that
of the $\Lambda $CDM model for both of the datasets. Let us here emphasize
that this enhancement in the value of $h$ for QSSC model, is because of its
oscillatory nature.

\begin{equation}
\alpha = 0.154^{+0.032}_{-0.028} \,, \quad \beta = 0.760^{+0.011}_{-0.011}
\,, \quad \eta = 0.366^{+0.091}_{-0.077} \,, \quad h =
0.718^{+0.018}_{-0.019} \,.  \label{best-fit}
\end{equation}

\begin{figure}[tbp]
\centering
\includegraphics[scale=1]{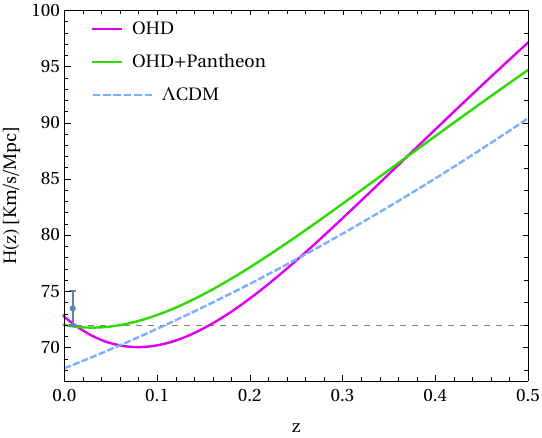}
\caption{Evolutionary profiles of $H(z)$ with $z \in[0,0.5]$ for OHD and its
combination with Pantheon dataset.}
\label{fig:H}
\end{figure}

With the above best-fit estimations, we plot the evolutionary behavior of $%
H(z)$ with $z$ in fig.\thinspace (\ref{fig:H}). In that figure, one can see
that $H(z)$ naturally tends to increase towards the current epoch which is
compatible with the low-redshift Hubble observations such as Riess et al. 
\cite{Riess}. This enhancement in $h$ not just come out naturally in QSSC
scenarios, but it also explain the large value of the Hubble constant from
the low-$z$ observations. From above estimate on $h$ we find that QSSC
significantly reduces the Hubble tension between high and low redshift
observations. In particular, we find that our estimate on $h$ is well within 
$1\sigma $ level with low-$z$ observations which removes the existing
tension on the Hubble constant. \textbf{Moreover, This model converges to }$%
\Lambda $\textbf{CDM in the past. Deviations from it appear only around the
present epoch, where Hubble starts increasing, mimicking a late time phantom
phase. Thereby, the model is as consistent with CMB as }$\Lambda $\textbf{%
CDM.}

Let us note that the model under consideration has no one to one
correspondence with the standard DE models, which directly take into account
the amount of matter and DE densities. Since, estimating the respective
densities is difficult in this case, therefore, in order to estimate the
corresponding DE equation of state parameter $w_{_{DE}}$, we consider Planck
2018 results best-fit for DE matter density parameter $\Omega_{_{DE}}$.

From Eqs. (\ref{w_param}) and (\ref{wtotal}) and using the best-fits given
in (\ref{best-fit}), one can estimate $w_{_{DE}}$ and the present age of the
universe $t(0)$ from Eq. (\ref{t_0}) as 
\begin{equation}
t(0)=45.05~Gyr\,,\quad w_{_{DE}}(0)\in \lbrack -1.036,-1.011]\,.
\label{wdecomb}
\end{equation}%
Here note that the large value of the present time is consistent with the
lower limit on the age of the universe i.e. $13.8$\thinspace Gyr \cite{age1,
age2, age3, age4}. This is due to the fact that in QSSC scenarios the
universe instead of having a Big-Bang event keep on oscillating, which as a
result gives rise to large value to the age of the universe. Moreover, the
age of the universe is large enough to be consistent with the age of the
globular clusters. Also note that within $1\sigma $, $w_{_{DE}}(0)$ shows
significant deviation from the $\Lambda $CDM. This implies that $%
w_{_{DE}}(0) $ has crossed the mark of $-1$ (corresponding to the $\Lambda $%
CDM) in the past and is currently in the phantom region. It is interesting
in a sense that this happens without incorporating any kind of assumption to
the DE component, but is completely stem out of the oscillatory nature of
the universe.

{\bf As we have shown above in fig.\,(\ref{fig:H}) that $H$ tends to be larger at the present epoch for the QSSC scenario, it is still intriguing 
or rather crucial to know whether this enhancement in $H$ or the ease of tension between the late-time and early-time cosmological data will still be preserved if one does not marginalize the absolute magnitude 
$M$. This is because in SH0ES actually measures the absolute magnitude 
$M$ of the type 1a Supernovae which are assumed to be standard 
candles by modulating Supernovae host galaxy distances to local
geometric distance anchor via the relation of Cepheid period 
luminosity. Then using the magnitude-redshift relation of the Pantheon 
sample, the value of $H_0$ is calculated from the measurement of $M$. 
As a result, it has been hypothesized that the $H_0$ tension is really a tension on the supernova absolute magnitude $M$, since the SH0ES $H_0$ measurement is based on $M$ estimations.  
	
Keeping in mind that the Hubble constant and absolute magnitude $M$ are highly correlated with one another, we 
now make a complete parametric estimation using the Pantheon and Hubble 
datasets on both $H_0$ and $M$. Our estimated parameters are enlisted below:
\begin{equation}
\alpha = 0.163^{+0.022}_{-0.020} \,, \quad \beta = 0.749^{+0.024}_{-0.017}
	\,, \quad \eta = 0.358^{+0.062}_{-0.047} \,, \quad h =
	0.719^{+0.047}_{-0.049} \,, \quad M =-19.302^{+0.047}_{-0.061} \label{best-fit2}
\end{equation}
}
\begin{figure}[!t]
	\begin{subfigure}{.49\textwidth} 
		\centering
		\includegraphics[height=6.8cm, width=7.4cm]{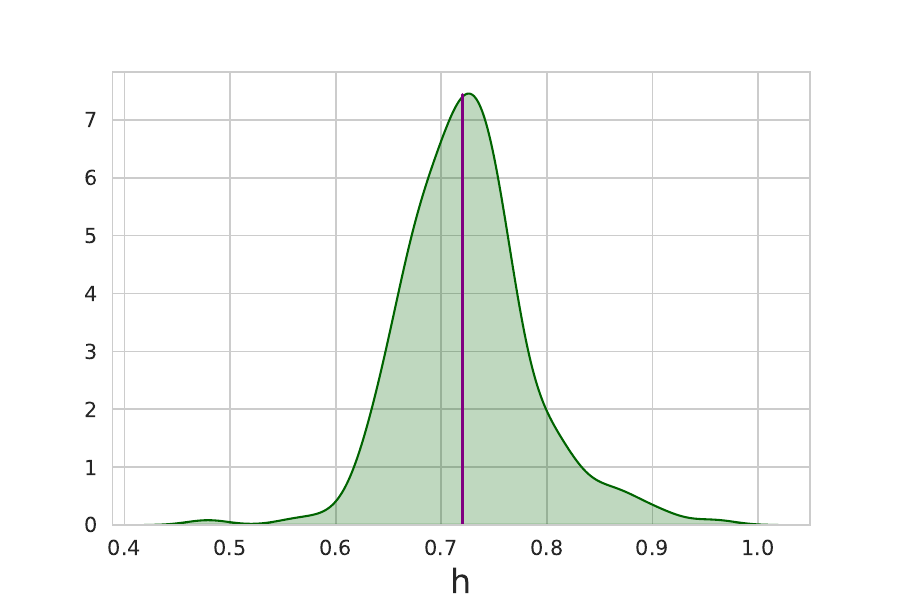}  
		\label{fig:Ha}
		\caption{Probability distribution of $h$.}
	\end{subfigure}
	\begin{subfigure}{.5\textwidth}
		\centering
		\includegraphics[height=6.8cm, width=7.4cm]{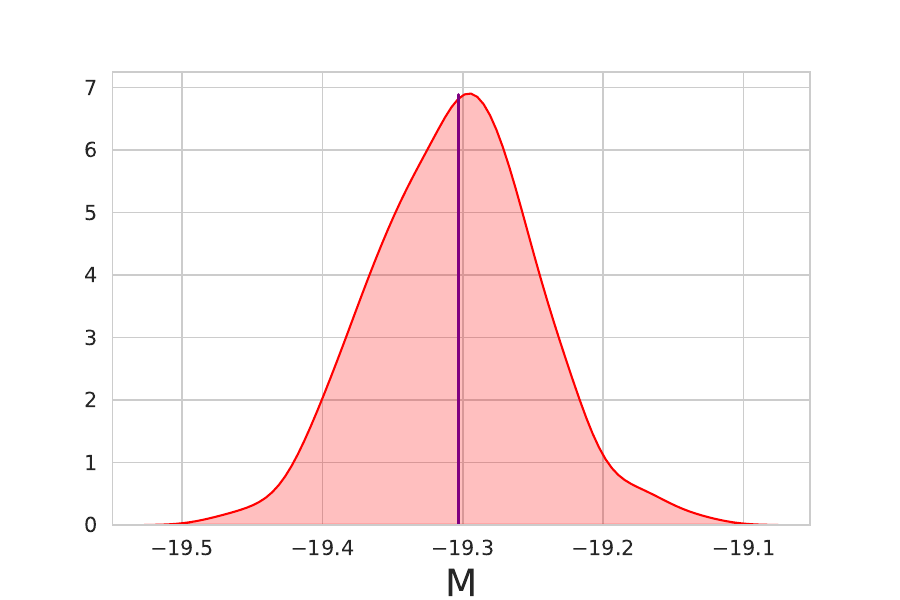}
		\label{fig:M}  
		\caption{Probability distribution of $M$.}
	\end{subfigure}
	\caption{Probability distribution functions for Hubble constant $h$ and 
		absolute magnitude $M$. The vertical line corresponds to their corresponding best-fit values.}
	\label{fig:HM}
\end{figure}
{\bf In above estimations, one can see that even with the supernovae absolute magnitude the value for the Hubble constant comes out to be almost the 
	same as we have in our previous estimations (\ref{best-fit}). It shows that 
	the QSSC model indeed predicts the enhancement of the Hubble parameter 
	near the present epoch due to the oscillatory nature of its scale factor.
	Also, note that our estimation on $M$ i.e. $-19.302^{+0.047}_{-0.061}$ 
	does not reflect any tension with its observed measurement. 
}

\section{Conclusion}

\label{conclusion}

In this paper, we have investigated the observational aspects of the QSSC
with the possibility of resolving the Hubble tension. In principle, QSSC
model involves an oscillatory behavior of the scale factor and hence the
Hubble parameter. The Hubble parameter, by nature of oscillatory, gets turn
around by reaching its minima near the present epoch (fig\,(\ref{fig:H})).
As a result, the model naturally exhibits super-acceleration expansion of
the Universe due to the enhancement in the Hubble parameter which become
compatible with low-redshift galaxy surveys.

One of the major requirements for the QSSC is to also be compatible with the
age of the Universe. Due to its oscillatory representation of the universe,
the cycle per oscillation must be greater than or equal to the age of the
Universe, otherwise it may contradict the age of the some of the local
galaxy clusters. In fact in our parametric estimations, we have explicitly
shown that the age of the Universe is not just satisfies the lower bound $%
13.8$\,Gyr but is even greater than $30$\,Gyr which is because of the
repeatedly expansions and contractions of the Universe. Notably, this is
also consistent with the age of the globular clusters and the Planck
constraints. We have also found that if there could be an effective DE
equation of state $w_{DE}$ associated with QSSC model, which if we can also
extract out from the total equation of state, which is further depend on the
Hubble parameter, then that effective DE behaves as a phantom at the,
atleast, present epoch. This is, however, interesting in a sense that there
the phantom nature of effective DE appears naturally without the
incorporation of any adhoc degree of freedom.

\section*{Acknowledgments}

We thank Prof. J. V. Narlikar, Emeritus Professor, IUCAA, Pune, Prof. M.
Sami, Director, CCSP and Dr. Mayukh Raj Gangopadhyay for fruitful
discussions, comments and suggestions to carry out the analysis in this
paper. One of the Author S. K. J. Pacif thank the Inter University Centre for Astronomy and Astrophysics (IUCAA) for hospitality and facility, where a part of the work has been carried out during a visit.

\end{document}